\documentclass[usenatbib]{mn2e}

\usepackage{epsfig}
\usepackage{mathrsfs,aas_macros}

\def\Lya{Ly$\alpha$~}

\def\HeII{\hbox{He$\,\rm \scriptstyle II\ $}}

\title[Resolving the \Lya forest in SPH simulations]{Resolving the high redshift
  Lyman-$\alpha$ forest in smoothed particle hydrodynamics
  simulations}

\author[J.S. Bolton \& G.D. Becker] {James S.
  Bolton$^{1}$ \& George D. Becker$^{2}$\\
  $^1$ Max Planck Institut f{\"u}r Astrophysik, Karl-Schwarzschild
  Str. 1, 85748 Garching, Germany \\
  $^2$  Kavli Institute for Cosmology and Institute of Astronomy, Madingley Road, Cambridge,
  CB3 0HA\\}

\begin{document}

\date{17 June 2009}

\maketitle

\label{firstpage}

\begin{abstract}

We use a large set of cosmological smoothed particle hydrodynamics
(SPH) simulations to examine the effect of mass resolution and box
size on synthetic \Lya forest spectra at $2\leq z \leq 5$.  The mass
resolution requirements for the convergence of the mean \Lya flux and
flux power spectrum at $z=5$ are significantly stricter than at lower
redshift.  This is because transmission in the high redshift \Lya
forest is primarily due to underdense regions in the intergalactic
medium (IGM), and these are less well resolved compared to the
moderately overdense regions which dominate the \Lya forest opacity at
$z\simeq 2-3$.  We further find that the gas density distribution in
our simulations differs significantly from previous results in the
literature at large overdensities ($\Delta>10$).  We conclude that
studies of the \Lya forest at $z=5$ using SPH simulations require a
gas particle mass of $M_{\rm gas}\leq 2\times 10^{5}\rm\,M_{\odot}/h$,
which is $\ga 8$ times the value required at $z=2$.  A box size of at
least $40\rm\,Mpc/h$ is preferable at all redshifts.

\end{abstract}
 


\section{Introduction}

Most observations of the \Lya forest are at $2 \leq z \leq 4$; this is
where the highest quality optical spectra and the largest data sets
are available.  Consequently, many studies of hydrodynamical \Lya
forest simulations also focus on this redshift range (for a review see
\citealt{Meiksin07rv}).  The convergence of a variety of simulated
\Lya forest statistics with resolution and box size has been explored
in detail at $z<4$, both for Eulerian grid and Lagrangian SPH
simulations (\citealt{Theuns98,Bryan99,Regan07}).  In recent years,
however, the importance of the $z\ga 4$ \Lya forest as a probe of the
hydrogen reionisation epoch (\citealt{Fan06,Becker07}) has led to
considerable interest in the properties of hydrodynamical \Lya forest
simulations at the {\it highest} observable redshifts
(\citealt{PaschosNorman05,BoltonHaehnelt07b}).

However, it is not obvious that the numerical requirements for
simulating the \Lya forest at $z\ga 4$ are the same as at lower
redshifts. The average transmission of the \Lya forest decreases
towards higher redshift as the physical gas density increases and the
intensity of the ultraviolet (UV) background falls.  Absorption lines
associated with the mildly overdense regions of the IGM at $z\simeq 2$
are gradually replaced by transmission peaks from underdense regions,
culminating in completely saturated absorption by $z\simeq 6$.  The
decrease in the characteristic overdensity associated with the \Lya
forest towards higher redshift will place strong demands on SPH
simulations, which naturally focus on resolving high density regions,
as a function of time.

In this letter we analyse a large set of cosmological simulations
performed with an upgraded version of the SPH code {\small GADGET-2}
(\citealt{Springel05}).  We consider the convergence of two widely
used \Lya forest flux statistics, the mean flux and the flux power
spectrum, with mass resolution and box size.  We also examine the gas
density distribution in our simulations in detail.  Although some of
the issues discussed in this letter are generally appreciated, there
have been no studies of SPH \Lya forest simulations at $z\simeq 5$
where they are highlighted explicitly.  Our intention is that these
results will provide a useful reference for future modelling
of the \Lya forest at the highest observable redshifts.


\vspace*{-0.6cm}
\section{Simulations}

We perform $24$ hydrodynamical simulations using {\small GADGET-3}, an
upgraded version of the publicly available parallel Tree-SPH code
{\small GADGET-2} (\citealt{Springel05}).  The simulations are
summarised in Table~\ref{tab:sims}, and cover a wide range of comoving
box sizes and gas particle masses.  The simulations are all started at
$z=99$, with initial conditions constructed using the same random
seed.  The cosmological parameters are ($\Omega_{\rm
  m},\Omega_{\Lambda},\Omega_{\rm b}h^{2},h,\sigma_{8},n_{\rm
  s})=(0.26,0.74,0.024,0.72,0.85,0.95)$.  These are consistent with
the fifth year {\it Wilkinson Microwave Anisotropy Probe} (WMAP) data
(\citealt{Dunkley09}) aside from $\sigma_{8}$, which is in better
agreement with \Lya forest constraints (\citealt{Viel04b,Seljak05}).
The gas is assumed to be of primordial composition with a helium mass
fraction of $Y=0.24$. The gravitational softening length is set to
$1/30^{\rm th}$ of the mean linear interparticle spacing.  Star
formation is included using a simplified prescription which converts
all gas particles with overdensity $\Delta=\rho/\langle \rho \rangle >
10^{3}$ and temperature $T<10^{5}\rm\,K$ into collisionless stars.

The baryons in the simulations are photoionised and heated by the UV
background model of \cite{HaardtMadau01} which includes emission from
both quasars and galaxies.  The UV background is switched on at $z=9$
and is applied in the optically thin limit using a non-equilibrium
ionisation algorithm.  The \HeII photo-heating rate is increased by a
factor of $1.8$ to give temperatures similar to existing measurements
(\citealt{Schaye00}).  Synthetic \Lya forest spectra are constructed
from each simulation at $z=(2,3,4,5)$.  Note that in this work we do
not subsequently alter the resolution or S/N of the spectra.

\begin{table}
\centering
  \caption{Mass resolution and box size (comoving) of the
    hydrodynamical simulations used in this work.}

  \begin{tabular}{c|c|c|c|c|c|c|c}
    \hline
    \hline
    Name     & $L$     & Total particle         & $M_{\rm gas}$  \\
             & [Mpc/h]   &  number  &  $[\rm M_{\odot}/h]$ \\
  \hline

  2.5-50     & 2.5    & $2 \times 50^{3}$    & $1.61 \times 10^{6}$ \\
  2.5-100    & 2.5    & $2 \times 100^{3}$   & $2.01 \times 10^{5}$ \\
  2.5-200    & 2.5    & $2 \times 200^{3}$   & $2.51 \times 10^{4}$ \\
  2.5-400    & 2.5    & $2 \times 400^{3}$   & $3.14 \times 10^{3}$ \\
  5-50       & 5      & $2 \times 50^{3}$    & $1.29 \times 10^{7}$ \\
  5-100      & 5      & $2 \times 100^{3}$   & $1.61 \times 10^{6}$ \\
  5-200      & 5      & $2 \times 200^{3}$   & $2.01 \times 10^{5}$ \\
  5-400      & 5      & $2 \times 400^{3}$   & $2.51 \times 10^{4}$ \\
  10-50      & 10     & $2 \times 50^{3}$    & $1.03 \times 10^{8}$ \\
  10-100     & 10     & $2 \times 100^{3}$   & $1.29 \times 10^{7}$ \\
  10-200     & 10     & $2 \times 200^{3}$   & $1.61 \times 10^{6}$ \\
  10-400     & 10     & $2 \times 400^{3}$   & $2.01 \times 10^{5}$ \\
  20-50      & 20     & $2 \times 50^{3}$    & $8.22 \times 10^{8}$ \\
  20-100     & 20     & $2 \times 100^{3}$   & $1.03 \times 10^{8}$ \\
  20-200     & 20     & $2 \times 200^{3}$   & $1.29 \times 10^{7}$ \\
  20-400     & 20     & $2 \times 400^{3}$   & $1.61 \times 10^{6}$ \\
  40-50      & 40     & $2 \times 50^{3}$    & $6.58 \times 10^{9}$ \\
  40-100     & 40     & $2 \times 100^{3}$   & $8.22 \times 10^{8}$ \\
  40-200     & 40     & $2 \times 200^{3}$   & $1.03 \times 10^{8}$ \\
  40-400     & 40     & $2 \times 400^{3}$   & $1.29 \times 10^{7}$ \\
  80-50      & 80     & $2 \times 50^{3}$    & $5.26 \times 10^{10}$ \\
  80-100     & 80     & $2 \times 100^{3}$   & $6.58 \times 10^{9}$ \\
  80-200     & 80     & $2 \times 200^{3}$   & $8.22 \times 10^{8}$ \\
  80-400     & 80     & $2 \times 400^{3}$   & $1.03 \times 10^{8}$ \\
   \hline
\end{tabular}
\label{tab:sims}
\end{table}


\vspace*{-0.5cm}

\section{Results}
\subsection{Convergence of flux statistics}

\begin{figure}
\begin{center}
  \includegraphics[width=0.45\textwidth]{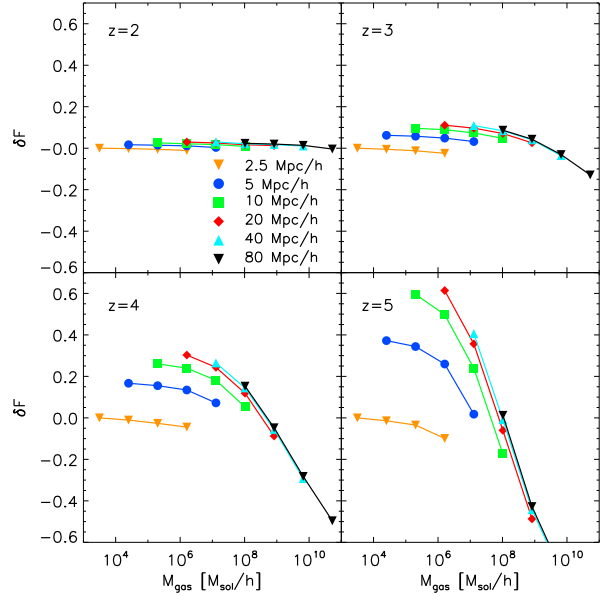}
\vspace{-0.1cm}
\caption{The difference in the mean \Lya flux, $\delta F = (\langle F
  \rangle - \langle F \rangle_{\rm f})/ \langle F \rangle_{\rm f}$
  relative to a fiducial value, $\langle F \rangle_{\rm f}$, as a
  function of gas particle mass for the hydrodynamical simulations
  listed in Table~\ref{tab:sims}.  The four panels display the results
  at $z=(2,3,4,5)$ where $\langle F \rangle_{\rm
    f}=(0.843,0.601,0.286,0.072)$, corresponding to the values in the
  highest resolution simulation (2.5-400).  Note, however, this is not
  necessarily the preferred model due to its small box size.  The
  relative difference becomes significantly larger at higher redshifts
  for decreasing mass resolution and box size.}
\label{fig:flux}
\end{center}
\end{figure}

We first consider the simplest \Lya forest observable, the mean flux,
$\langle F \rangle = \langle e^{-\tau}\rangle$, in
Fig.~\ref{fig:flux}.  This shows the difference in $\langle F \rangle$
measured from our synthetic spectra relative to the highest resolution
model (2.5-400) as a function of gas particle mass at $z=(2,3,4,5)$.
At $z=2$ the mean flux converges quickly with mass resolution. There
is a 0.5 (1.3) per cent different between the 20-200 (20-100) and
20-400 models, giving a good degree convergence with resolution for
$M_{\rm gas}=1.61\times 10^{6}\rm \,M_{\odot}/h$.  Convergence of
$\langle F \rangle$ with box size may be judged by the vertical
separation between data points with the same gas particle mass.  A
$20$ Mpc/h box provides acceptable results at $z=2$, with a 0.7 (0.2)
percent difference between the 20-100 (40-200) and 80-400 models.
\cite{Tytler09}, who recently examined the convergence of $\langle F
\rangle$ with box size at $z=2$ using the Eulerian hydrodynamical code
Enzo, find similar results.

However, it is clear that the relative difference becomes
significantly larger with increasing redshift and decreasing mass
resolution.  By $z=5$, there is a 2.1 (8.2) per cent difference
between $\langle F \rangle $ in the 5-200 (5-100) and 5-400 models.
Only a marginal degree of convergence with mass resolution is achieved
for $M_{\rm gas}=2.01\times 10^{5}\rm \,M_{\odot}/h$.  The differences
due to box size also become larger, with a 7.2 (2.1) per cent
difference between $\langle F \rangle$ in the 20-100 (40-200) and
80-400 models.  Clearly, the box size and mass resolution requirements
for simulating the mean flux of the \Lya forest are much stricter at
higher redshift.

\begin{figure}
\begin{center}
  \includegraphics[width=0.45\textwidth]{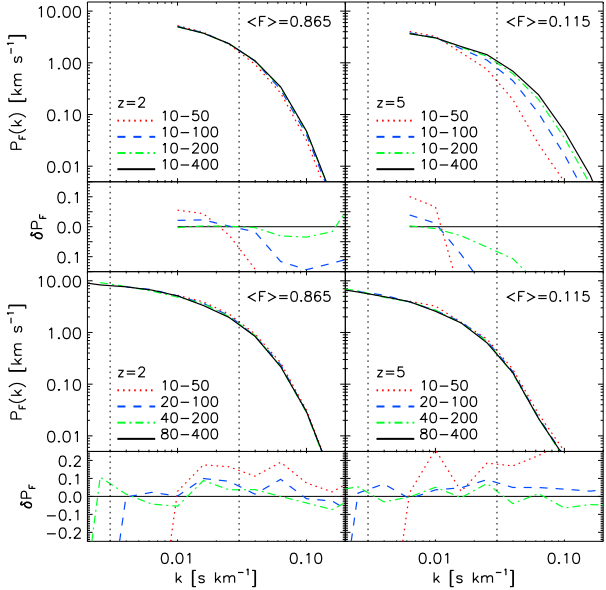}
\vspace{-0.2cm}
\caption{The upper panels show the 1D flux power spectrum computed
  from simulations with fixed box size ($10\rm\,Mpc/h$) at $z=2$ and
  $z=5$, while the lower panels display the power spectrum from
  simulations with fixed mass resolution ($1.03 \times
  10^{8}\rm\,M_{\odot}/h$).  The spectra have been rescaled to have
  the same mean flux, indicated in the upper right of each panel.  The
  difference in the power spectrum relative to the models represented
  by the solid curves is shown in the lower third of each panel.  The
  vertical dotted lines bracket the range of wavenumbers used by
  \citet{Viel04b} to infer the amplitude and shape of the matter power
  spectrum, $0.003<k\,[{\rm s\,km^{-1}}]<0.03$.}
\label{fig:power}
\end{center}
\end{figure}

The second statistic we consider is the 1D \Lya flux power spectrum.
This has been extensively used as a probe of the primordial matter
power spectrum on scales of $0.5-40\rm\,Mpc/h$ at $2\leq z \leq 4$,
and there are several studies which examine its convergence with
resolution and box size in some detail (\citealt{McDonald03,Viel04b}).
There has been comparatively little work performed at higher redshifts
(but see \citealt{Viel08}).  The upper panels in Fig.~\ref{fig:power}
display the power spectrum of the \Lya flux, $F=e^{-\tau}$, at $z=2$
and $z=5$ computed from the simulations with a box size of 10 Mpc/h.
The vertical dotted lines bracket the range of wavenumbers used by
\cite{Viel04b} to infer the amplitude and shape of the matter power
spectrum at $z<3$.  Note that we have rescaled the synthetic spectra
to have the same $\langle F \rangle$ for this comparison.  The
convergence with mass resolution is again significantly poorer at
$z=5$ and is also scale dependent.  For the range $0.003<k\,[{\rm
    s\,km^{-1}}]<0.03$ at $z=2$ the 10-200 (10-100) data is within $1$
(3) per cent of the 10-400 model, corresponding to a requirement for
$M_{\rm gas}=1.61\times 10^{6}\rm\,M_{\odot}/h$.  However, at $z=5$,
the 10-200 (10-100) model deviates from 10-400 by up to 7 (22) per
cent, while the 5-200 (5-100) simulations (not shown) deviate from
5-400 by 2 (6) per cent. A mass resolution of at least $M_{\rm
  gas}=2.01\times 10^{5}\rm\,M_{\odot}/h$ is required at $z=5$ for
marginal convergence.

The lower panels in Fig.~\ref{fig:power} display the effect of box
size on $P_{\rm F}(k)$ for fixed mass resolution.  At $z=5$, the
20-100 (40-200) model is within 10 (8) per cent of the 80-400 model,
while at $z=2$ the 20-100 (40-200) model is within 10 (9) per cent of
the 80-400 model.  A box size of at least $40\rm\,Mpc/h$ is therefore
required to adequately model the power spectrum at $2\leq z \leq 5$.
Similar requirements have been noted by \cite{McDonald03} and
 \cite{Viel04b} at $z\simeq 2-3$.  We briefly note we also performed an
examination of a third flux statistic, the distribution of the \Lya
flux.  We again found the simulation requirements to be significantly
stricter towards higher redshifts, with the same box size and
resolution requirements as the mean flux.  We do not report these
results in detail here.

\subsection{The gas density distribution}

\begin{figure}
\begin{center}
  \includegraphics[width=0.45\textwidth]{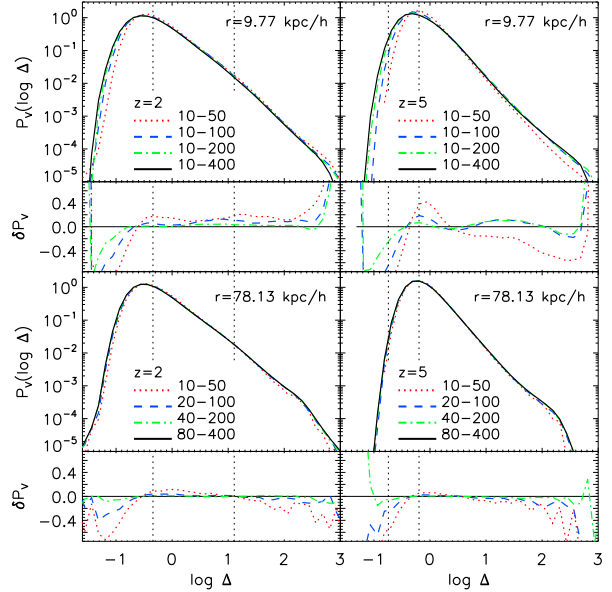}
\vspace{-0.2cm}
\caption{The upper panels show the volume weighted gas density
  distribution computed from simulations with fixed box size
  ($10\rm\,Mpc/h$) at $z=2$ and $z=5$, while the lower panels display
  the density distribution from simulations with fixed mass resolution
  ($1.03 \times 10^{8}\rm\,M_{\odot}/h$). The difference in the
  density distribution relative to the models represented by the solid
  curves is shown in the lower third of each panel.  The vertical
  dotted lines in each panel bracket the range of optical depth
  weighted overdensities corresponding to 95 per cent of all the
  pixels with $0.05 \leq F \leq 0.95$ in the associated \Lya forest
  spectra.}
\label{fig:Dpdf}
\end{center}
\end{figure}

The explanation for these results is apparent on inspecting the gas
density distribution in the simulations.  The upper panels in
Fig.~\ref{fig:Dpdf} display the volume weighted density distribution
per unit $\log \Delta$ at $z=2$ and $z=5$ for models with a box size
of $10\rm\,Mpc/h$, while the lower panels display the results for
fixed mass resolution.  The density distributions are obtained by
interpolating the particle masses, weighted by the smoothing kernel,
onto a regular grid with a cell size, $r$, indicated in each panel.
The vertical dotted lines in each panel bracket the range of optical
depth weighted overdensities corresponding to 95 per cent of all the
pixels with $0.05 \leq F \leq 0.95$ in the associated \Lya forest
spectra.  The mean flux of the spectra at $z=(2,5)$ is $\langle F
\rangle=(0.865,0.115)$.

At $z=2$ the 10-200 (10-100) density distribution is within 5 (13) per
cent of the 10-400 model for $-0.5 \le \log \Delta \le 2.5$, while at
$z=5$ the 10-200 (10-100) is within 12 (18) per cent of 10-400,
providing at best marginal convergence.  Outwith this density range
the distribution has not converged at either redshift.  In
the lower two panels the 40-200 (20-100) is within 3 (12) per cent of
80-400 at $z=2$, and 7 (18) per cent at $z=5$.  It is the poor
convergence with mass resolution and box size in the most underdense
regions ($\log \Delta<-0.5$) which drives the \Lya forest results.  At
$z=2$, the \Lya forest is dominated by gas with $\Delta>1$.  The
transmission from underdense regions is always close to the continuum
($F=1$) regardless of mass resolution, and so under-resolving these
regions has little impact on the \Lya flux statistics.  In contrast,
underdense regions dominate the transmission at $z=5$, impacting
significantly on the convergence of the simulated \Lya forest properties.

\begin{table*}
\centering
\begin{minipage}{180mm}
\begin{center}

  \caption{Tabulated coefficients for the eighth order polynomial fits
    to the gas density distribution from our 10-400 simulation,
    $\log[P_{\rm V}(x)]=\sum_{i}a_{i}x^{i}$, where $x=\log
    \Delta$. $P_{V}(\Delta)$ is normalised to unity and $\Delta P_{\rm
      V}(\Delta)=P(\log \Delta)/\ln 10$. Note that the fits are made
    over the range $-1 \leq \log \Delta \leq 2.5$ {\it only}, and are
    all within $<5$ per cent of the simulation data for $-0.5 \leq
    \log \Delta \leq 2.5$.}
  \begin{tabular}{c|c|c|c|c|c|c|c|c|c}
    \hline
    \hline
    $z$     & $a_{0}$ & $a_{1}$ & $a_{2}$ & $a_{3}$ & $a_{4}$ &
    $a_{5}$ & $a_{6}$ & $a_{7}$ & $a_{8}$  \\
    \hline

7.0 &   -0.038744
    &  -1.193136
    &  -1.209168
    &   1.480778
    &  -1.355202
    &  0.649847
   & -0.093190
   & -0.024297
   & 0.006604
\\

6.5 & -0.045600
    &  -1.162781
    &  -1.187344
    &   1.312205
    &  -1.225248
    &  0.647885
    & -0.120218
   & -0.014682
   & 0.005503
\\

6.0 &-0.059247
    &  -1.148617
    &  -1.122149
    &   1.176996
    &  -1.178106
    &  0.661365
    & -0.105707
    &-0.031420
    &0.008956
\\

5.5 &-0.077189
    &  -1.115055
    &  -1.049649
    &  0.913821
    &  -1.018321
    &  0.774770
    & -0.259515
    & 0.025978
    & 0.001650
\\

5.0  & -0.089459
     & -1.083824
     & -1.051325
     & 0.669231
    & -0.709611
    &  0.777568
    & -0.414747
    &  0.101665
   &-0.009361
\\

4.5 & -0.102984
    &  -1.100786
    &  -1.064773
    &  0.677866
   &  -0.542385
   &   0.556266
   &  -0.298218
   &  0.072713
   &-0.006496
\\

4.0 & -0.132602
    &  -1.132429
    & -0.970648
    &  0.719472
    & -0.552924
    &  0.418740
    & -0.169566
    & 0.029132
   &-0.001301
\\

3.5 &   -0.163770
    &  -1.134785
    & -0.885486
    &  0.611691
    & -0.404073
    &  0.374816
    & -0.230005
    & 0.070230
   &-0.008295
\\

3.0 &-0.203625
    &  -1.166205
    & -0.784556
    &  0.642483
    & -0.324878
    &  0.199731
    & -0.147317
    & 0.059290
  & -0.008639
\\

2.5&   -0.259451
   &   -1.190049
   &  -0.614461
   &   0.661790
   &  -0.367005
   &  0.090334
   & -0.034647
   &  0.021446
   & -0.004244
\\

2.0 &-0.325057
    &  -1.184408
    & -0.442799
    &  0.566567
    & -0.347644
    & 0.074473
    &-0.016128
    & 0.012290
   &-0.002695

\\

\hline
\end{tabular}
\end{center}
\end{minipage}
\label{tab:fits}
\end{table*}

\begin{figure}
\begin{center}
  \includegraphics[width=0.45\textwidth]{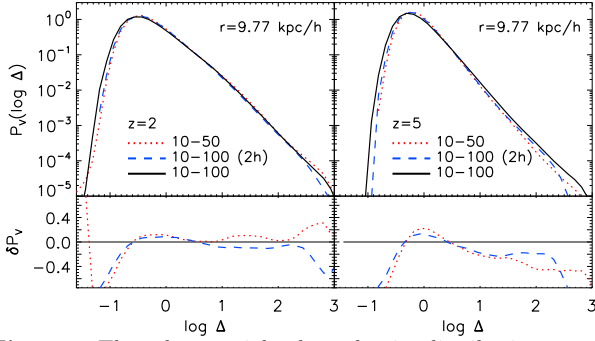}
\vspace{-0.3cm}
\caption{The volume weighted gas density distribution at $z=2$ (left
  panel) and $z=5$ (right panel) from the 10-50 (dotted curve) and
  10-100 (solid curve) models.  The dashed curve is computed from the
  10-100 simulation after doubling the particle smoothing lengths,
  matching those used in the 10-50 model in order to mimic lower mass
  resolution.}
\label{fig:SPHtest}
\end{center}
\end{figure}

This behaviour is a consequence of the spatially adaptive nature of
SPH, which provides excellent spatial resolution in high density
regions but poorer resolution in underdense regions.  In
Fig.~\ref{fig:SPHtest} we demonstrate this by comparing the density
distribution from the 10-50 and 10-100 models to a third distribution,
again drawn from the 10-100 model.  The latter is computed by
interpolating the particle masses onto a grid using the 10-50 model
particle smoothing lengths (twice the 10-100 values), mimicking
particle masses a factor of eight larger.  The 10-50 distribution at
low densities is well reproduced by the resmoothed 10-100
distribution, implying that differences in the density distribution at
$\log \Delta<-0.5$ are largely a consequence of the intrinsic mass
resolution limit.  However, additional effects such as gas being
transferred from the low density IGM to previously unresolved haloes
may also play a small role (\citealt{Theuns98}).

\begin{figure}
\begin{center}
  \includegraphics[width=0.45\textwidth]{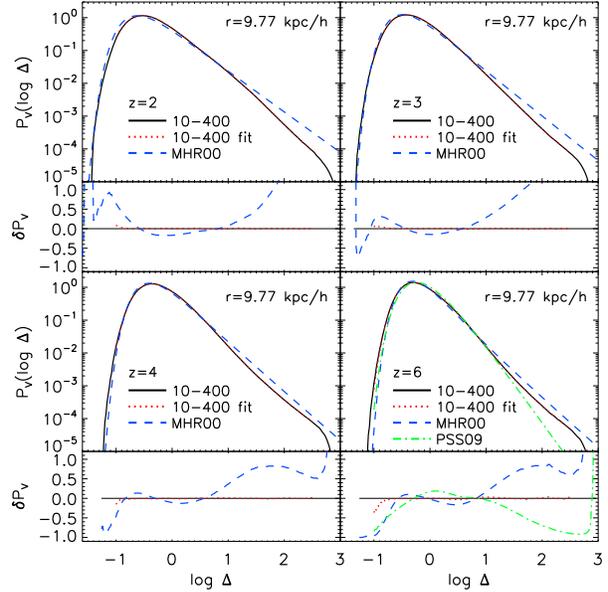}
\vspace{-0.2cm}
\caption{The volume weighted gas density distribution extracted from
  the 10-400 simulation at $z=(2,3,4,6)$ (solid curves).  The dotted
  curves correspond to an eighth order polynomial fit to the
  simulation data over the range $-1\leq \log \Delta \leq 2.5$.  The
  fits obtained by MHR00 correspond to the dashed curves.  The
  dot-dashed curve in the panel at $z=6$ also shows the recent fit
  presented by PSS09.  The differences in the fits relative to the
  simulation data are shown in the lower third of each panel.}
\label{fig:fit}
\end{center}
\end{figure}

Lastly, we consider fits to the gas density distribution which are
widely used in analytical models of the \Lya forest.  In
Fig.~\ref{fig:fit} we compare the volume weighted density distribution
from our 10-400 simulation to the four parameter fits\footnote{The
  MHR00 fits are derived from the L10 simulation of
  \cite{MiraldaEscude96}, which uses ($\Omega_{\rm
    m},\Omega_{\Lambda},\Omega_{\rm b}h^{2},h,\sigma_{8},n_{\rm
    s})=(0.4,0.6,0.015,0.65,0.79,0.96)$, with a box size of
  $10\rm\,Mpc/h$ box and $288^{3}$ cells.  This gives an average gas
  mass per cell of $6\times 10^{5}\rm\,M_{\odot}$.  Note the $z=6$ MHR00
  distribution is an extrapolation from the lower redshift fits.}
obtained at $z=(2,3,4)$ by \cite{MiraldaEscude00} (hereafter MHR00).
The solid curves in Fig.~\ref{fig:fit} correspond to our 10-400
simulation data, while the dashed curves show the fits obtained by
MHR00.  The dot-dashed curve at $z=6$ corresponds to a more recent
fit\footnote{PSS09 use a \small{GADGET-2} simulation with
  ($\Omega_{\rm m},\Omega_{\Lambda},\Omega_{\rm
    b}h^{2},h,\sigma_{8},n_{\rm
    s})=(0.258,0.742,0.0228,0.719,0.796,0.963)$, a box size of
  $6.25\rm\,Mpc/h$ and $M_{\rm gas}=1.8\times 10^{5}\rm\,M_{\odot}/h$
  ($256^{3}$ gas particles).}  to an SPH simulation by \cite{Pawlik09}
(hereafter PSS09), who also use the parameterisation suggested by
MHR00.

Although our simulation is in reasonable agreement with the MHR00 fits
for $-0.5\leq \log \Delta \leq 1$, we confirm the claim by PSS09 that
the power law tail in the MHR00 parameterisation, $P_{\rm
  V}(\Delta)\propto \Delta^{-\beta}$ for $\Delta \gg 1$, provides a
poor description of the density distribution.  This is perhaps not too
surprising; the MHR00 prescription is based on the assumption of a
power-law density profile for collapsed objects and is not obtained
directly from the simulations.  The 10-400 model also differs
considerably from the MHR00 fits at $\log \Delta<-0.5$, although note
that our data are not fully converged here.  The PSS09 fit is in poor
agreement with our simulation at $z=6$.  However, PSS09 find a similar
discrepancy between their fit and simulation results, suggesting that
the difference between their {\it simulated} density distribution and
our data is actually much smaller.  Furthermore, we use a very
different star formation prescription to PSS09 which is designed to
optimise \Lya forest simulations.  We have verified this has little
effect on the simulated gas density distribution at $\log\Delta<2$
when compared to a more sophisticated star formation prescription
(\citealt{Springel03}), but this choice will produce differences in
the density distribution at higher densities.  

We conclude that the MHR00 parameterisation is not fully adequate for
describing $P_{\rm V}$ from our simulations at $\log\Delta>1$.
Consequently, we provide polynomial fits to the 10-400 density
distribution in Table 2 over the range $-1\leq \log \Delta \leq 2.5$
{\it only} (dotted curves in Fig.~\ref{fig:fit}).  We deliberately
avoid parameterising the data, preferring to instead provide an
accurate representation of the simulations for reference.  Note,
however, that $P_{\rm V}$ is still only marginally converged with
resolution for $-0.5 \leq \log \Delta \leq 2.5$ and has not fully
converged with box size.


\vspace*{-0.5cm}

\section{Conclusions}

We perform a large set of cosmological SPH simulations to explore the
effect of mass resolution and box size on two key \Lya forest flux
statistics.  As noted in many previous studies (see
\citealt{Meiksin07rv} for a review), we find a mass resolution of
$M_{\rm gas}=1.61\times 10^{6}\rm \,M_{\odot}/h$ is more than adequate
for simulating mean \Lya flux and power spectrum at $z=2$. However,
towards higher redshift the mass resolution requirement for
convergence becomes significantly stricter, with a gas particle mass
{\it at least} 8 times smaller required at $z=5$.  We demonstrate this
is largely a consequence of the intrinsic resolution limit of SPH
simulations in low density regions.  Although a $20\rm\,Mpc/h$ box is
adequate for simulating the mean flux at $z=2$, a $40\rm\,Mpc/h$ box
is required for the power spectrum, and this is preferred for both
statistics at $z=5$.  We also briefly demonstrate that the MHR00
parameterisation for the gas density distribution, although in
reasonable agreement with our 10-400 model at moderate overdensities,
provides a poor description of the simulation for $\Delta>10$.

Although our results will hold in general for SPH \Lya forest
simulations, in detail they will only be exact for the specific models
we present.  Convergence requirements will always depend on the
physical process under consideration, as well as the precision of the
observational data to which the simulations are compared.  We have
also assumed that the $z\ga 5$ forest is dominated by transmission
from underdense regions in the IGM.  This picture is correct if the UV
background is spatially uniform, but if there are large fluctuations
in the ionising radiation field at $z\ga 5$, localised patches of
highly ionised gas will complicate this interpretation somewhat.
Lastly, at lower redshifts where the \Lya forest transmission is
dominated by mildly overdense regions, SPH simulations produce results
comparable to those from state-of-the-art adaptive mesh codes
(\citealt{Regan07}).  It would be very interesting to perform a
similar comparison at $z\simeq 5$.

\vspace*{-0.5cm}

\section*{Acknowledgements}

We thank M.G. Haehnelt, M. McQuinn, S.P. Oh, J. Pritchard, V. Springel
and M. Viel for valuable comments.  The simulations used in this work
were performed using the COSMOS facility at DAMTP in Cambridge. COSMOS
is sponsored by SGI, Intel, HEFCE and STFC.  GDB acknowledges
financial support from the Kavli foundation.

\end{document}